\documentclass[11pt,a4paper]{article}
\usepackage{amsmath, amssymb}
\usepackage{natbib}
\usepackage{graphicx}
\usepackage[hmargin=2.5cm]{geometry}
\usepackage{lineno}

\author{Nicolas \textsc{Brantut}$^{1,2}$, Fran\c cois X. \textsc{Passel\`egue}$^3$, Pierre \textsc{Dublanchet}$^4$}
\title{Pore pressure change during nucleation and slip along experimental faults}
\date{$^1$ GFZ Helmholtz Centre for Geosciences, Potsdam, Germany.\\$^2$ Department of Earth Sciences, University College London, London, UK.\\$^3$Universit\'e C\^ote d'Azur, CNRS, Observatoire de la C\^ote d'Azur, IRD, G\'eoazur, Sophia Antipolis, France.\\$^4$Mines Paris, PSL University, Centre for Geosciences and Geoengineering, Fontainebleau, France.}

\begin{document}

\maketitle

\begin{abstract}
  Fluid pressure within faults is coupled to fault slip, mechanical deformation of the fault walls, and fault frictional strength. In order to clarify the main controlling factors influencing fluid pressure changes in fault zones during the seismic cycle, we conducted laboratory rock friction experiments where fluid pressure was monitored in situ during sequences of quasi-static loading followed by dynamic slip events. The simulated fault was a $30^\circ$ saw-cut in a Westerly granite cylinder, saturated with water, tested under triaxial conditions. Pore pressure was held constant at the boundaries of the block, but the low hydraulic diffusivity of Westerly granite made the fault hydraulically disconnected from the boundaries. During quasi-static loading while the fault was locked, we observed pore pressure increases which we interpret as poroelastic closure of the fault. During dynamic slip events, pore pressure systematically dropped by amplitudes commensurate to the normal stress drop (shear and normal stress being coupled in triaxial conditions). A large contribution to the pore pressure drop is interpreted as poroelastic opening of the fault. Deviations from the poroelastic effects are observed: in small events, pore pressure dropped further than anticipated, indicating inelastic dilation. In a few large events, pore pressure dropped less than anticipated, which could be the sign of compaction or thermal pressurisation. Prior to macroscopic slip events, we detect systematic pore pressure decreases by up to around $1$~MPa, correlated to the occurrence of inhomogeneous preslip along the fault. Slip nucleation, inferred by kinematic inversion of local strain gauge data, is linked to local slip magnitudes of the order of $1$ to $10$~\textmu{}m, and appears to lead to inelastic dilation. A stability analysis of fault slip including dilatant and poroelastic effects shows that poroelastic coupling tends to compensate normal stress variations, leading to faults operating under mostly constant effective normal stress if conditions are undrained. Our experiments show the concurrent operation of poroelastic and inelastic fault closure/opening, with inelastic dilation detectable at the early stages of slip, and poroelastic effects dominating during dynamic slip when normal stress changes rapidly. 
\end{abstract}

Highlights:
\begin{itemize}
\item We measure experimentally on-fault pore pressure changes during slip on granite surfaces under trixial conditions,
\item Pre-slip is marked by small pore pressure drops, interpreted as inelastic dilatancy occurring in the first few \textmu{}m of slip,
\item One major contribution to pore pressure change during stick-slip in triaxial conditions is poroelastic decompression linked to normal stress drop,
\item Poroelastic pore pressure changes counteract normal stress changes, which impact fault stability.
\end{itemize}

\section{Introduction}


During the seismic cycle, pore fluid pressure in and around fault zones is expected to change due to compaction and dilation induced by deformation.  The bulk material surrounding faults is typically impacted by poroelastic effects in response to static and dynamic stress changes produced by earthquake or creep deformation. Such poroelastic pore pressure changes lead to ground deformation that can be measured by geodetic techniques \citep[e.g.][]{peltzer98,jonsson03}. By contrast, pore pressure perturbations due to pore volume change induced by slip or damage localised within or close to fault cores do not necessarily have direct detectable signatures in geophysical data. Quantifying these local changes is however crucial to our understanding of fault mechanics, because fault strength is to first order proportional to the (local) fault zone fluid pressure via the effective stress principle \citep[][Sec. 1.1]{scholz19}. 

Laboratory rock deformation experiments have shown that, in initially intact low porosity rocks such as granite, bulk inelastic deformation, fault formation and slip are linked to large relative pore volume increases \citep[e.g.][]{brace66b,rummel78}, which lead to pore pressure drops when deformation occurs under undrained or partially undrained conditions, i.e., when pore volume change occurs much faster than fluid diffusion to a distant reservoir \citep{brantut20}. Such pore pressure drops have important consequences for the dynamics of faulting: they can lead to hardening \citep[e.g.][]{brace68}, rupture stabilisation \citep{aben21}, and to the generation of rapid afterslip due to postseismic fluid pressure reequilibration \citep{aben23}.


Friction experiments conducted in simulated faults with planar geometry (saw cuts with bare rock surfaces or artificial gouge), representative of ``mature'' fault systems, have shown a more complex behaviour. Gouge dilatancy is typically reported in response to velocity steps \citep{morrow89,marone90,samuelson09,ashman23}, and associated pore pressure drops were inferred under undrained conditions \citep{lockner94}, but slip accumulation tends to produce long-term compaction and potential pore pressure rise \citep[e.g.][]{faulkner18}. Direct in-situ pore pressure measurements during friction experiments in saw cut granite and quartz gouge conducted under macroscopically undrained conditions \citep{proctor20} show a complex interplay between slip-induced dilation that lead to pore pressure drops, and compaction that leads to pore pressure increases and promote slip. This behaviour does not appear to be easily reconciled with conventional friction models based on rate-and-state friction \citep{segall95}. The laboratory work of \citet{proctor20} has thus revealed a major gap in our understanding of the hydro-mechanical behaviour of faults: we are now at a stage where it is difficult to make predictions regarding the spontaneous sense of variation and amplitude of pore pressure changes during fault slip.

Here, we aim to bring further direct constraints on this problem, specifically focussing on the variations in pore pressure prior to and during stick-slip events on bare rock surfaces. \citet{proctor20} showed that some large stick-slip events in bare rock surfaces can be preceded by compaction and pore pressure increase, while slow slip events were associated with pre-event dilation and pore pressure drops. The amplitude of the pore pressure changes reported by \citet{proctor20} were of the order of 0.1 to 1~MPa. What remains to be confirmed is how systematic these variations are over multiple events at different loading rates and effective pressure, and how closely linked they are to slip at the interface.

We conducted triaxial friction experiments on saw-cut Westerly granite under water-saturated conditions, with measurements of in-situ fluid pressure and local strain. We used local strain measurements to detect the nucleation phase of slip events \citep[e.g.][]{dublanchet24}, and correlate the timing and amplitude of preslip to on-fault pressure changes. In the triaxial configuration used here, we find that on-fault pore pressure changes are dominated by poroelastic effects, with a fault zone pore pressure that responds primarily to normal stress changes. Superimposed on this poroelastic effect, we detect systematic dilatant pore pressure drops that coincide with the preslip phase. The coseismic phase is found to be largely dominated by the poroelastic effect.


\section{Methods}


\subsection{Sample material and instrumentation}

\begin{figure}
  \centering
  \includegraphics{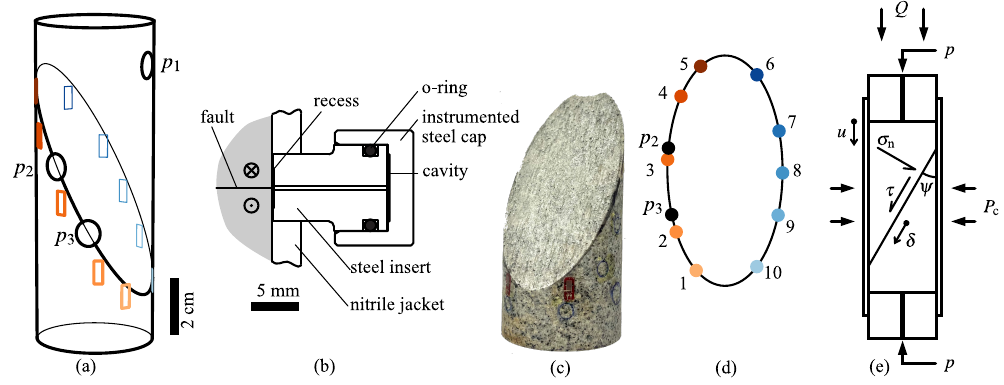}
  \caption{Sample and sensor geometry. (a) Schematic of the sample assembly, showing the location of pore pressure sensors $p_1$, $p_2$ and $p_3$, and axial strain gauges (red and blue rectangles). (b) Schematic of the pore pressure sensor in contact with the fault zone. The jacket is sealed around the stem with epoxy. (c) Photograph of the lower-half sample (taken after deformation). (d) Map view of the fault surface, with projected positions of strain gauges and pore pressure sensors. (e) Schematic of the loading geometry.}
  \label{fig:samplesketch}
\end{figure}

We used cylindrical cores of Westerly Granite, 40~mm in diameter and 100~mm in length, sawn at a 30$^\circ$ angle to the cylinder axis. The saw-cut surfaces were ground flat to ensure parallelism of the fault walls, and then roughened with \#220 SiC powder. An array of 10 linear strain gauges was positioned along the fault (Figure \ref{fig:samplesketch}). Each strain gauge (TML FLAB-3-350-11, gauge length 3~mm) was oriented along the axial direction, and bonded to the rock surface a few mm away from the fault. The sample was then inserted in a nitrile sleeve that was perforated to accommodate three local pore pressure sensors \citep[see details in][]{brantut20,brantut21}. One sensor was located in the off-fault region, near the sample end (labelled $p_1$ in Figure \ref{fig:samplesketch}), and two were located directly on the fault (labelled $p_2$ and $p_3$). In order to ensure good hydraulic connectivity between the fault and the internal space of the pore pressure sensor, a 200~\textmu m recess was machined at the end of the steel inserts in contact with the sample, and the inserts were carefully positioned across the fault (Figure \ref{fig:samplesketch}b). The additional dead volume is of the order of 2.5~mm$^3$, which brings the total dead volume of the transducers to a value of $5.4$~mm$^3$ \citep{brantut21}.

The deformation experiments were performed in the triaxial rock physics ensemble at University College London, originally described by \citet{clint00}. The jacketed sample was inserted between stainless steel endcaps and placed in the pressure vessel. Confining pressure ($P_\mathrm{c}$)was raised with an electric pump, controlled to an accuracy of 0.4~MPa, and measured at the inlet of the pressure vessel with a pressure transducer (precision of 0.01~MPa). Axial load was applied with a servohydraulic ram via an internally compensated piston. Sample shortening was measured by averaging the displacement measured externally by three linear variable differential transformers (LVDTs), and corrected for machine stiffness (480~kN/mm). Each sample end cap was hydraulically connected to an independent servohydraulic fluid intensifier. The fluid pressure at both ends of the sample was measured externally using pressure transducers (precision 0.01~MPa), and intensifier fluid volume was measured using a displacement transducer attached to the intensifier piston. The frequency of all servo-control feedback loops was 200~Hz.

Conditioned signals from strain gauges, pressure transducers, load cell and LVDTs were recorded at 1~Hz. In addition, digital oscilloscopes were used to record a subset of the signals (strain gauge, internal pore pressure, axial shortening) at 20~kHz during rapid deformation events. The oscilloscope recordings were started manually during each loading stage, data were streamed continuously and stopped after slip events. The high frequency recordings were smoothed using a moving average with a window of 200 samples.

\subsection{Calibrations and data processing}

The signals from internal pore pressure sensors were calibrated using the technique given in \citet{brantut21}: we imposed a range of constant, uniform confining and pore pressures, and recorded output voltages after stabilisation of the signals. Calibration factors (pressure to voltage) were determined by linear regression, and pore pressure was computed based on sensor voltage and the known confining pressure (assumed uniform across the sample) at all times during the test. We observed an irregular drift of the local pore pressure sensors by up to 1~MPa for sensors $p_1$ (off-fault) and $p_2$ (on-fault), and by up to 0.5~MPa for sensor $p_3$ (on-fault) over the 100~h duration of the test, which provides a limit for the absolute accuracy achieved in these measurements.

The load-point displacement was first corrected for the machine stiffness to obtain the sample shortening. Then, the sample shortening was corrected for the sample's stiffness to obtain the net slip during deformation. For each loading stage, we used linear fit of the shortening vs. stress data during the initial elastic loading phase to extract the sample stiffness and subtract the elastic contribution to the shortening.

The differential stress, confining pressure and slip data were used to determine the shear and normal stress on the fault, using the technique described by \citet{tembe10} which accounts for fault surface area decrease with increasing slip. The friction coefficient was computed as the ratio of resolved shear stress over Terzaghi's effective normal stress, using the local pore pressure data \citep[as in][]{proctor20}.

\subsection{Initial characterisation and deformation procedure}

The initially dry sample was first pressurised to a confining pressure of 20~MPa, with the pore fluid system left dry at atmospheric pressure. After stabilisation of pressure and strains, an upstream pore water pressure of 15~MPa was set and the downstream end of the sample was left vented to the atmosphere. The sample was left to saturate in this fashion for 7.5~hours, after which we observed a stabilisation of the internal pore pressure data. Using the pressure differential between the upstream and sensor $p_1$ (around 2~MPa) and the measured fluid flow rate ($1.74\times10^{-11}$~m$^3$/s), we estimated a permeability of the wall rock of the order of $8.8\times10^{-20}$~m$^2$. The average permeability across the entire sample was $9.2\times10^{-20}$~m$^2$, which is consistent with the fact that the average permeability is limited by the low-permeability fault walls, independently of the possibly large along-fault permeability.

After this initial pressurisation and saturation step, the downstream and upstream pore pressure lines were connected together to equilibrate fluid pressure to 15~MPa throughout the sample. During this phase, we observed a gradual homogenisation of the internal pore pressure, over a characteristic timescale of the order of 1000~s. This timescale is consistent with a hydraulic diffusivity of the order of $10^{-5}$~m$^2$/s. After full stabilisation of all pressures and strains, we conducted a series of confining pressure and pore pressure steps, waiting each time for full stabilisation of the signals, in order to calibrate the internal pressure sensors (see above). During each step, we systematically observed a timescale of the order of 1000~s to achieve full stabilisation, consistent with our initial diffusivity estimate and with previously published data on a similar sample \citep[e.g.][]{brantut21}.

After all calibration steps were completed, we set the confining pressure and pore pressures to the target nominal values (initially, 80~MPa and 20~MPa, respectively), and started deformation at constant load-point displacement rate (initially, 0.2~\textmu m/s). After a ``run-in'' stage of around 0.6~mm (stable) net slip (around 1~mm load point displacement), we unloaded the sample, and subsequently conducted a series of loading stages where we imposed constant load-point displacement rate until the onset of slip, and then unloaded the sample. The full set of conditions tested are reported in Table \ref{tab:conditions}, and the full time series is shown in Supplementary Figure S1.

\begin{table}
  \centering
  \caption{Conditions of all deformation stages and slip events. All reported values were extracted from low frequency recordings. The reported pore pressure is that imposed at the sample boundaries, and equilibrated prior to loading. Shear, normal stress and fault slip were calculated from differential stress, confining pressure and corrected shortening data.}
  \label{tab:conditions}
  \begin{tabular}{cccccccc}
    \hline
    & Confining  & Pore & Loading & Shear & Normal& Pore pressure & \\
    & pressure   & pressure & rate & stress drop & stress drop & drop ($\Delta p_2,\Delta p_3$) & Slip \\
    Stage & (MPa) & (MPa) & (\textmu m/s) & (MPa) & (MPa) & (MPa) & (mm) \\
    \hline
    Run-in & 80 & 20 & 0.2 &  &  & &   \\
    1 & 80 & 20 & 1.0 & 6.81 & 3.8 & 2.93, 4.24 & 0.073 \\
    2 & 80 & 20 & 1.0 & 3.53 & 1.96 & 1.8, 2.59 & 0.038 \\
    3 & 80 & 20 & 5.1 & 29.69 & 16.63 & 17.13, 20.2 & 0.32 \\
    4 & 80 & 20 & 5.0 & 26.47 & 14.76 & 17.55, 20.27 & 0.299 \\
    5 & 80 & 20 & 2.5 & 27.17 & 15.17 & 16.84, 19.46 & 0.298 \\
    6 & 79 & 20 & 2.5 & 26.46 & 14.75 & 16.03, 18.94 & 0.297 \\
    7 & 79 & 50 & 1.0 & 5.82 & 3.27 & 3.04, 3.89 & 0.063 \\
    8 & 79 & 50 & 2.5 & 2.51 & 1.43 & 1.68, 2.0 & 0.031 \\
    9 & 80 & 50 & 5.0 & 2.08 & 1.14 & 1.61, 1.92 & 0.043 \\
    10 & 80 & 50 & 1.0 & 1.78 & 0.97 & 1.27, 1.41 & 0.021 \\
    11 & 80 & 35 & 1.0 & 0.63 & 0.34 & 1.55, 1.52 & 0.014 \\
    12 & 80 & 35 & 2.5 & 2.84 & 1.55 & 1.7, 2.39 & 0.032 \\
    13 & 80 & 35 & 5.0 & 4.43 & 2.47 & 2.54, 3.51 & 0.052 \\
    14 & 80 & 35 & 5.0 & 2.76 & 1.56 & 2.48, 3.39 & 0.041 \\
    Run-in & 110 & 20 & 0.2 &  & & &   \\
    15 & 110 & 20 & 1.0 & 7.05 & 3.86 & 3.22, 4.62 & 0.073 \\
    16 & 110 & 20 & 2.5 & 5.55 & 3.05 & 3.2, 4.53 & 0.06 \\
    17 & 110 & 20 & 5.0 & 6.69 & 3.68 & 3.72, 5.23 & 0.071 \\
    18 & 110 & 20 & 5.0 & 62.61 & 34.51 & 18.1, 21.1 & 0.664 \\
    19 & 110 & 20 & 5.0 & 46.52 & 25.53 & 18.12, 23.89 & 0.547 \\
    20 & 110 & 20 & 1.0 & 48.38 & 26.57 & 14.38, 23.57 & 0.538 \\
    21 & 139 & 20 & 1.0 & 4.6 & 2.55 & 1.31, 2.44 & 0.044 \\
    \hline
  \end{tabular}
\end{table}

\subsection{Slip inversion}

For selected slip events, the local strain data prior to macroscopic slip was analysed to obtain estimates of local slip heterogeneity along the fault, using the method described in \citet{dublanchet24}. In this method, the sample is assumed to be elastic and isotropic, subject to constant displacement boundary conditions at each end and constant confining pressure on its cylindrical surface. Inhomogeneous slip on the fault leads to elastic strains in the bulk. The strains at each strain gauge position can then be used to infer which region of the fault slipped, and by how much.

The fault is discretised into 27 nodes where a scalar slip magnitude is defined (slip is only assumed to occur along the dip direction of the fault). The forward model is computed using the appropriate elasto-static Green's functions accounting for the cylindrical geometry of the sample, and the experimental boundary conditions (obtained with finite element approach). The data are the axial strains recorded at each strain gauge location. The high frequency (20~kHz), smoothed (200 points moving average) record of local strain gauges was first corrected for a linear trend to extract only variations with respect to the overall increasing elastic load. This dataset was then first inverted to obtain local slip maps using a damped least-square method (termed deterministic inversion step in \citet{dublanchet24}), leading to smooth mean models. These mean models were then used as starting points for systematic sampling of the posterior probability distribution of slip, using a Markov Chain Monte Carlo method.

The overall outcome of the method is a sequence of slip maps, with the statistical distribution of slip values at each fault node. Here, we will not analyse in detail each inversion result, but only extract key qualitative features of the inverted slip maps.

\section{Results}


\subsection{General features}

\begin{figure}
  \centering
  \includegraphics{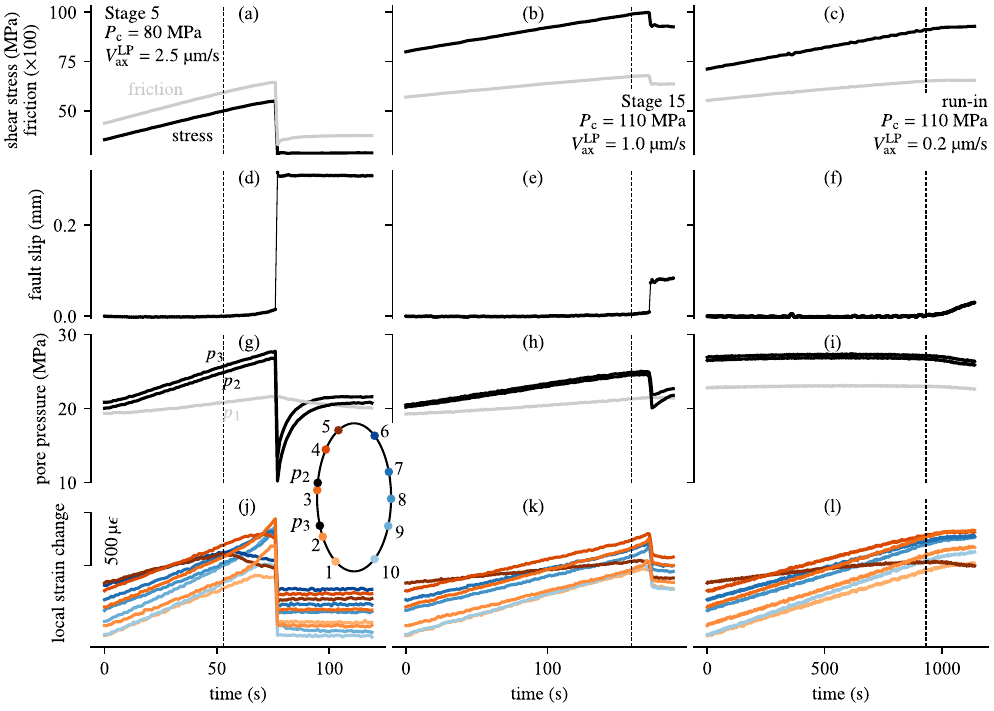}
  \caption{Representative examples of the shear stress, friction coefficient, pore pressure and slip evolution during loading at different load-point velocities ($V_\mathrm{LP}$), confining and initial pore pressure. The local strain data are offset by the along-strike position of each strain gauge, and colour-coded according to gauge position as in Figure \ref{fig:samplesketch}. Vertical dashed line show the onset of non-uniform fault slip, as picked from the divergence of the strain gauge records from a linear trend.}
  \label{fig:examples}
\end{figure}

\begin{figure}
  \centering
  \includegraphics{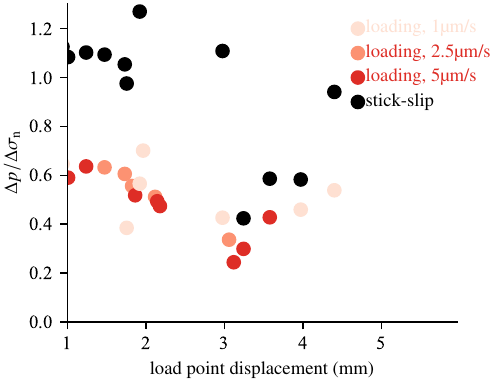}
  \caption{Evolution of the ratio $\Delta p/\Delta\sigma_\mathrm{n}$ during the loading stage (red dots) and coseismic unloading stage (black dots) as a function of total slip, using the record from sensor 3 (Figure \protect\ref{fig:samplesketch}). During the loading stage, the ratio was computed by a linear fit of the pore pressure vs. normal stress record in the early stage of loading, at stresses well below the onset of slip. During coseismic unloading, the ratio was computed from the two data points recorded after and before the slip event.}
  \label{fig:dpload}
\end{figure}

In all tests, the pore pressure on and off the fault initially increased with increasing load prior to the onset of slip (Figure \ref{fig:examples}). The pore pressure increase was systematically larger on the fault than off the fault, and followed a linear trend with increasing fault normal stress $\sigma_\mathrm{n}$, with proportionality factor $\Delta p_\mathrm{f}/\Delta\sigma_\mathrm{n}$ ranging from $0.2$ to $0.7$ (Figure \ref{fig:dpload}). This factor did not exhibit any systematic trend with loading rate or initial effective pressure, but seems to decrease slightly with increasing total accumulated slip on the fault (Figure \ref{fig:dpload}).

In the case of the ``run-in'' phase, the onset of slip was gradual and linked with hardening (Figure \ref{fig:examples}). During this first phase, the pore pressure showed a slight decrease over time with ongoing slip. Concomitantly, the strain gauge array recorded a gradual change, with the development of stress heterogeneity linked to inhomogeneous slip distribution along the fault \citep[see Section \ref{sec:preslip};][]{dublanchet24}.

In all other cases, slip on the fault was marked by a macroscopic friction (and stress) drop, and was associated with a concomitant pore pressure drop. Prior to the macroscopic stress drop (recorded by the load cell), inhomogeneous fault slip could be detected by the divergence of the local strain records. The onset of inhomogeneous slip marks the beginning of the pre-slip phase, during which small but nonzero average slip can be detected from the corrected LVDT data. The macroscopic stress drop corresponds to the relative motion of the fault walls as (approximately) solid blocks, and is marked by a collective and near-simultaneous unloading of all local strain gauges. Following each slip event, no further deformation or slip could be detected, while the pore pressure gradually returned to a constant value, over a timescale of the order of 10~seconds.

\begin{figure}
  \centering
  \includegraphics{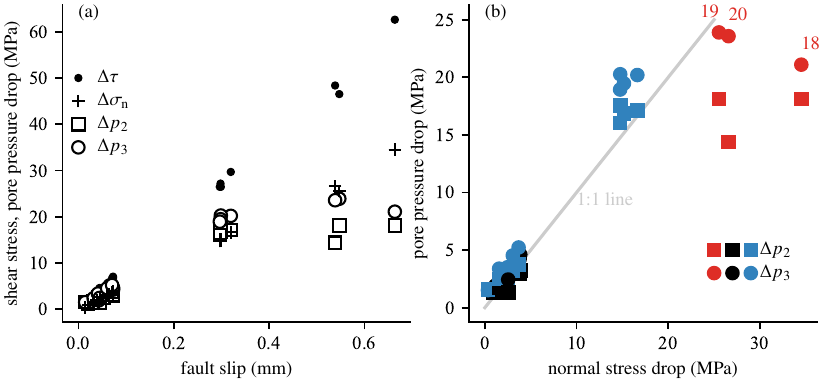}
  \caption{(a) Shear, normal stress and pore pressure drops as a function fo slip and (b) pore pressure drop as a function of  normal stress drop during all stick-slip events. Blue markers indicate pore pressure drops that are larger than 1~MPa above the normal stress drop, and red markers indicates pressure drops smaller than 1~MPa below normal stress drop. Numbers indicate event index as reported in Table \ref{tab:conditions}.}
  \label{fig:pdrop}
\end{figure}

The slip accumulated in each event was proportional to the shear stress drop (Figure \ref{fig:pdrop}), as expected for block motion in a spring-slider configuration \citep[e.g.][]{lockner17}. The apparent shear stiffness of the system (machine and sample) was 88~MPa/mm. For all slip events where slip was less than around 0.4~mm (i.e., stress drop less than around 40~MPa), the on-fault pore pressure drop was found to be directly proportional to the normal stress drop, with a ratio of $\Delta p_\mathrm{f}/\Delta \sigma_\mathrm{n}$ of around $1$ (Figure \ref{fig:pdrop}b, \ref{fig:dpload}). 

\begin{figure}
  \centering
  \includegraphics{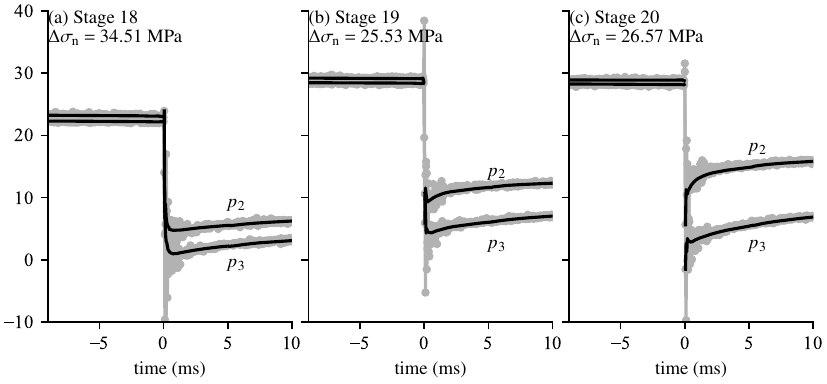}
  \caption{Details of the pore pressure change during the three largest events (stage numbers refer to Table \ref{tab:conditions}), where pore pressure drop was slightly lower than purely anticipated from poroelastic decompression. The grey symbols and lines show the calibrated output from the raw, unfiltered, $20$~kHz recordings of the pore pressure sensors. Black lines show moving averages of that data before and after the main drop, using a 200 point window.}
  \label{fig:largeslip}
\end{figure}

The three largest events, with slip ranging from 0.5 to 0.7~mm, were associated with moderate pore pressure drops, substantially smaller than the normal stress drops with a ratio $\Delta p_\mathrm{f}/\Delta \sigma_\mathrm{n}$ ranging from 0.4 to 0.6. In those events, the initial pore pressure at the beginning of loading was 20~MPa (see Table \ref{tab:conditions}), and by the onset of macroscopic slip the pore pressure had increased up to around 30~MPa. We measured maximum pressure drops between 22 and 30~MPa (Figure \ref{fig:largeslip}), i.e., the pore pressure nearly dropped to zero during slip.

\subsection{Preslip phase}
\label{sec:preslip}

\begin{figure}
  \centering
  \includegraphics{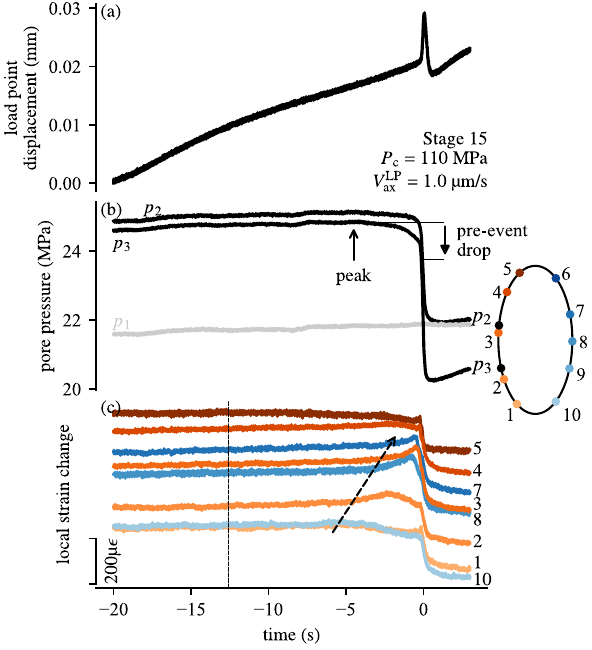}
  \caption{Example of high frequency record of load-point displacement, pore pressure and local strain as a function of time. The local strain data are offset by the along-strike position of each strain gauge, and colour-coded according to gauge position as in Figure \ref{fig:samplesketch}. Arrow highlights the location of the peak pore pressure during the preslip phase. Dashed arrow highlights the trend of slip patch growth as observed on the local strain data. The vertical dashed line shows the onset of departure from linearity of the strain record, here visible on gauge 5.}
  \label{fig:preslip_example}
\end{figure}

The period beginning at the onset of inhomogeneous slip (marked by the divergence of the local strain records) until the macroscopic stress drop typically lasted between a few seconds to a few tens of seconds. Figure \ref{fig:preslip_example} shows a representative example of this ``preslip'' phase. During preslip, the load-point displacement kept increasing linearly with time (tracking the servo-controlled set point), only showing a sharp oscillation at the time of the macroscopic stress drop due to transient loss of control \citep{lockner17}. In the few tens of seconds preceding the macroscopic stress drop, the pore pressure was initially constant (the long-term increase during elastic loading is not detectable over that timescale), reached a maximum (see arrow in Figure \ref{fig:preslip_example}) and started to decrease by a few bars until the rapid drop linked to the main stress drop. In the example shown in Figure \ref{fig:preslip_example}, the onset of local on-fault pore pressure decrease on each sensor ($p_2$ and $p_3$) coincides with the onset of local strain deviation near those locations (see gauges numbered 3 and 2 in Figure \ref{fig:samplesketch}d, respectively). The local strain change shows that deviations from the elastic trend initiate near the bottom end of the sample (gauges 1 and 10) and propagate slowly upward (propagation rate of the order of 10~mm/s, , see dashed arrow in Figure \ref{fig:preslip_example}), at which point the macroscopic stress drop occurs and all strain data abruptly drop.

\begin{figure}
  \centering
  \includegraphics{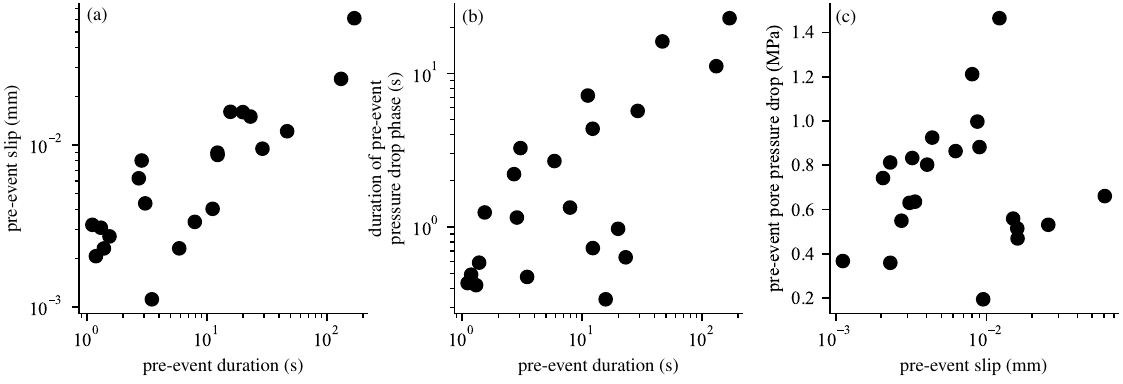}
  \caption{Correlations between preslip duration, amount of preslip, duration of pore pressure drop and amount of pore pressure drop.}
  \label{fig:preslip}
\end{figure}

We determined the duration of the preslip phase for all events, and tried to establish possible correlations between that duration and other quantities characterising preslip. The amount of preslip, as detected from the external shortening corrected for machine and sample stiffness, is seen to generally increase  with increasing preslip duration (Figure \ref{fig:preslip}a). The timing of the onset of pore pressure decrease is quite variable, but an increasing trend with increasing preslip duration can be detected (Figure \ref{fig:preslip}b). The amount of pore pressure drop prior to the macroscopic stress drop is not clearly correlated to other preslip characteristics, and only a weak trend towards increasing pore pressure drop with increasing preslip is detected (Figure \ref{fig:preslip}c).

\begin{figure}
  \centering
  \includegraphics{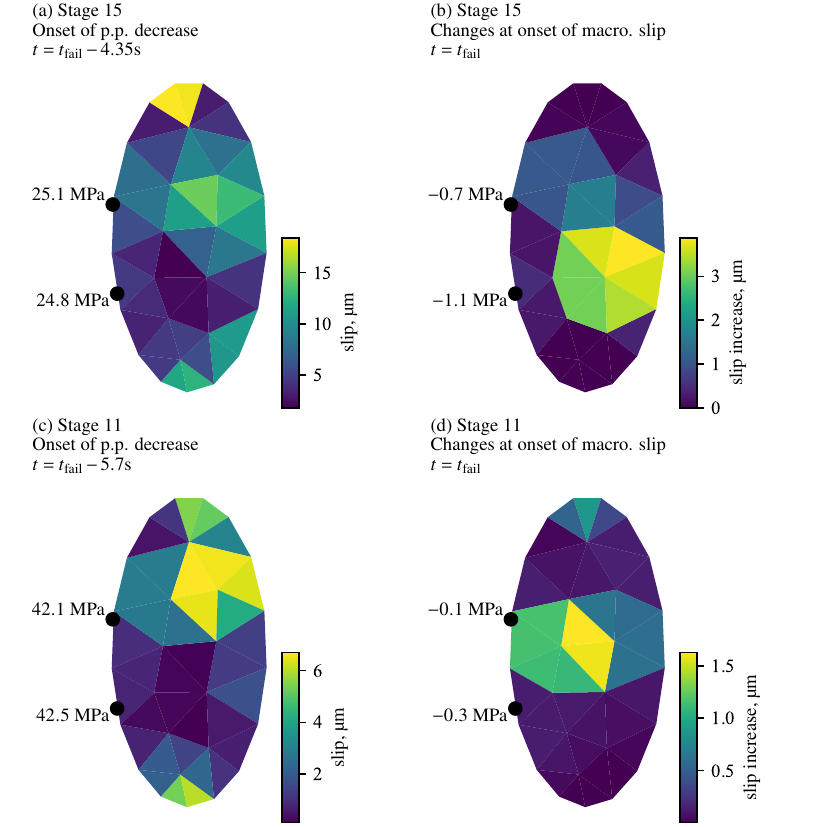}
  \caption{Examples of slip distribution at the onset of pore pressure decrease (a,c), and slip variation between that point and the onset of macroscopic slip (b,d) for two events (stages 10 and 15 in Table \ref{tab:conditions}). The black dots indicate the positions of pore pressure sensors and their absolute values (a,c) and variations before macroscopic slip (b,d).}
  \label{fig:slipmaps}
\end{figure}

The slip inversion provides more detail regarding the nucleation phase and its relationship with pore pressure variations. The onset of nucleation is characterised by heterogeneous slip along the fault, with slip magnitudes of up to 10s of \textmu{}m (see representative examples in Figure \ref{fig:slipmaps}a,c). During this early phase, no substantial deviation in pore pressure change can be detected on the fault. In the few seconds to 10s of seconds prior to the macroscopic stress drop, slip further accumulates locally by several \textmu{}m, that is, at slip rates of the order of \textmu{}m/s, and pore pressure locally decreases by up to around 1~MPa. As hinted above in the averaged out observations (Figure \ref{fig:preslip}), there is little correlation between the exact slip magnitude and location and the detectable magnitude of the pore pressure decrease.

\section{Discussion}

\subsection{Dynamic pore pressure drop: Elastic vs. inelastic effects} \label{sec:elastic}


The main result of our experiments is the observation of a systematic on-fault pore pressure drop concomitant with the macroscopic stress drop for all slip events. This result is qualitatively consistent with those of \citet{proctor20} for bare rock surfaces, and indicate that slip events are associated with fault dilation. The magnitude of the pressure drops observed here, up to around 20~MPa for slip events of around 0.5~mm, is typically larger than observed by \citet{proctor20}, but our experimental conditions were slightly different. Firstly, \citet{proctor20} only observed one large stick-slip event, followed by slower events with limited stress drops. Secondly, we used custom pore pressure sensors with a limited dead volume that was probably smaller than the dead volume in the configuration used by \citet{proctor20} (see potential additional space between fault and sensor in their Figure 1b).

In triaxial conditions, variations in axial load on an oblique saw-cut fault necessarily produce both shear and normal stress changes, so it is difficult to separate the potential dilatancy effect of slip itself from the poroelastic effects due to normal stress variations.

For events with slip less than 0.5~mm, the pore pressure drop is nearly equal to the normal stress drop (Figures \ref{fig:dpload}, \ref{fig:pdrop}b), which could be explained by the dominance of a poroelastic decompression effect with a ratio $\Delta p/\Delta\sigma_\mathrm{n}$ (equivalent to the normal Skempton coefficient) close to 1 for the fault zone material. Under undrained conditions, relevant for the very short timescales involved during stick slip, this ratio can be expressed as \citep[e.g.,][p. 382]{kachanov93}
\begin{equation}
  \Delta p/\Delta\sigma_\mathrm{n} = \frac{1}{1 + \beta_\mathrm{f} w/C_\mathrm{n}},
\end{equation}
where $\beta_\mathrm{f}$ is the compressibility of the fluid, $w$ is the fracture aperture and $C_\mathrm{n}$ is the normal elastic compliance of the fracture. A ratio $\Delta p/\Delta\sigma_\mathrm{n}$ close to one would imply that the normal compliance of the fracture is large compared to the compliance of the filling fluid. Fracture compliance is typically stress- and roughness-dependent \citep{jaeger07}, and it is difficult to make
accurate estimates of this quantity for our specific sample configuration (made even more complex by the presence of gouge generated by the successive slip events). However, it is well established that the normal compliance of open voids in an elastic matrix scales with $1/(E\zeta)$, where $E$ is the Young's modulus of the matrix and $\zeta$ is the void aspect ratio \citep[e.g.][]{jaeger07}. Thus, if we consider the macroscopic fracture to be well represented by an array of colinear voids  with low aspect ratio \citep{myer00}, its effective normal compliance should be very large, and the ratio $\Delta p/\Delta\sigma_\mathrm{n}$ would indeed be expected to be close to 1. Such theoretical considerations are consistent with direct measurements in fractured rock \citep{chapman24}.

If the poroelastic pore pressure change in the fault is nearly equal to the normal stress change, then deviations from that response indicate the occurrence of inelastic compaction, dilation or thermal effects. There is some scatter in the pore pressure drop data, but many events produced pore pressure drops that were larger than 1~MPa above the normal stress drop (blue markers in Figure \ref{fig:pdrop}b), which indicates that inelastic coseismic dilatancy likely contributed to the coseismic pore pressure drop, consistently with \citeauthor{proctor20}'s observations.

The three largest events with slip greater than 0.5~mm showed limited pore pressure drops, significantly smaller than predicted by a pure poroelastic decompression. However, for those events, the normal stress drop was greater than the initial pore pressure prior to stick-slip, and the pore pressure drops were nearly total (Figure \ref{fig:largeslip}). It is thus likely that the normal stress drops have produced sufficient poroelastic fluid decompression to reach the degassing or vaporisation pressure of the saturating fluid, which would naturally limit the pore pressure drop. The amount of slip in each event follows the trend expected from stick-slip dynamics \citep[proportional to the stress drop, see][; Figure \protect\ref{fig:pdrop}a]{shimamoto80}, and the larger stress drops compared to other events can be explained by the differences in initial effective stress, load-point velocity and hold times prior to loading (Table \ref{tab:conditions}), as well as spontaneous variability typical of triaxial stick-slip data \citep[e.g.][, their Figure 6.5]{lockner17}.  Therefore, the three large stick-slip events do not appear abnormal and it is difficult to conclude that they reflect any specific pressurisation process (compaction or thermal pressurisation).

We anticipate that slip-induced pore fluid pressurisation might only be detectable at larger slip (for the thermal pressurisation mechanism) or in unconsolidated/immature faults (for shear-induced compaction). Previous experimental work has shown evidence for pore fluid pressurisation in both contexts. In rotary shear experiments in water saturated granite, \citet{badt20} inferred that thermal pressurisation of pore fluid could be responsible for rapid weakening over slip displacements of 2 to 5 mm. \citet{aretusini21} tested artifical fault gouges at elevated slip rate, and reported pore fluid pressurisation by mechnical compaction. \citet{faulkner18} and \citet{proctor20} also reported compaction-driven pressurisation in slow slip events under triaxial conditions. In our experiments, the fault walls have low permeability and slip rate is expected to be high during dynamic slip events, so in principle the fault could be prone to thermal pressurisation. Theoretical predictions for the characteristic slip required for the activation of thermal pressurisation in nominally intact granite is indeed of the order of 1~mm \citep{rice06,badt20}. However, during stick-slip instabilities, \citet{segall06} report that thermal pressurisation only becomes significant at slip distances much larger than those associated with dilation and rate-and-state effects. Indeed, shear-induced dilation (elastic or inelastic) is likely dominant at small slip: our preslip observations show that pore pressure may drop by as much as 1~MPa in a few \textmu{}m. Thermal pressurisation can somewhat compensate dilation by producing more heat \citep{garagash03a}, but it is not likely to be effective if the characteristic slip distance associated with dilation is much smaller than that associated with thermal pressurisation \citep[][Appendix C]{brantut21}. Furthermore, potential shear-induced compaction effects may not be detectable in our experiments since we subjected the fault to around 1~mm slip in a run-in phase, after which wear and fault closure rates would already be limited \citep[e.g.][]{wang94}.


\subsection{Preslip dilatancy}

We systematically observed a pore pressure decrease in association with the onset of (partial) slip along the fault. The pore pressure decrease prior to macroscopic stress drop and slip can be up to around 1~MPa, for pre-slip amounts up to around 10~\textmu{}m. Slip inversion reveals that the pore pressure drops can be linked to slip magnitudes as low as a few \textmu{}m.

The preslip pore pressure decrease is not associated with any detectable normal stress variation, so it must correspond to inelastic dilatancy localised along the fault interface, i.e., fault zone opening. This behaviour is qualitatively consistent with the bare rock experiment reported by \citet{proctor20}, where a slow slip event was preceded by a decrease in pore pressure and accelerated preslip. More generally, precursory dilatancy is consistent with a range of laboratory observations in artificial gouge or bare rock surfaces, whereby macroscopic slip events (dynamic or slow) are preceded by drops in elastic wave speeds \citep[e.g.][]{tinti16,scuderi16} and local accelerated acoustic emission activity \citep[e.g.][]{thompson05,johnson13}. The magnitude of fault dilation is quantitatively minimal: For a 1~MPa pore pressure drop, the corresponding relative fault zone widening should be of the order of $5\times10^{-4}$ if undrained conditions are assumed (taking a fluid compressibility of $5\times10^{-10}$~Pa$^{-1}$). For a fault width of the order of 100~\textmu{}m, the opening would be of the order of 50~nm.
This small magnitude is on the same order as that measured in fault gouge and captured by the ``dilatancy coefficient'' defined by \citet[][their Equation 15]{segall95} as the porosity increase in response to an $e$-fold increase in slip rate. In quartz gouge, this dilatancy coefficient ranges from $0.5\times10^{-4}$ to $4\times10^{-4}$ \citep{segall95,samuelson09}, and may be as high as $10^{-3}$ in quartz-clay mixtures \citep{ashman23}. The dilatancy coefficient cannot be directly measured in our experimental configuration, but as illustrated above the associated pore pressure drop in the preslip phase is consistent with the order of magnitude documented in the literature.

It is noteworthy that preslip dilatancy did not seem to have any systematic impact on the eventual stability of slip: there is no apparent correlation between the magnitude of preslip pore pressure drop and slip velocity. The detection of a stabilising effect would require extensive characterisation of the frictional constitutive parameters and slip history, which was not attempted here.

\subsection{Comparison with intact rocks}

Our results on bare rock surfaces and planar fault geometry highlight the possibly dominant role of on-fault poroelastic processes during co-seismic slip. Such poroelastic effects were not accounted for in previous work conducted in initially intact rock \citep{brantut20,aben21,liu24}, where the on-fault pore pressure variations were considered to be dominated by inelastic dilational effects. The contribution of poroelastic decompression in the dramatic pore pressure drops observed during rock failure by \citet{brantut20,aben21} is difficult to assess, but is likely not dominant. Firstly, there is a substantial inelastic dilation during rupture, measured by pore volume change before/after failure, which must be negatively impacting the local pore pressure. Secondly, \citet{aben23} report pore pressure drops of 5 to 7~MPa in response to partial failure with modest shear stress drops of the order of 2~MPa (i.e., around 1~MPa normal stress drop on the prospective fault plane), which are inconsistent with a purely poroelastic response.

In rough faults, such as those produced by spontaneous failure of intact cylinders under triaxial conditions, the role of poroelastic pore pressure changes during slip events is likely not negligible, and we anticipate a more complex interaction between poroelastic and inelastic effects. The post-failure slip events reported by \citet{aben21} show poroelastic pore pressure increases during the loading stage, and slip is accompanied by pore pressure drops, in accordance with both elastic and inelastic dilation effects.


\subsection{Drainage state of the fault}

\begin{figure}
  \centering
  \includegraphics{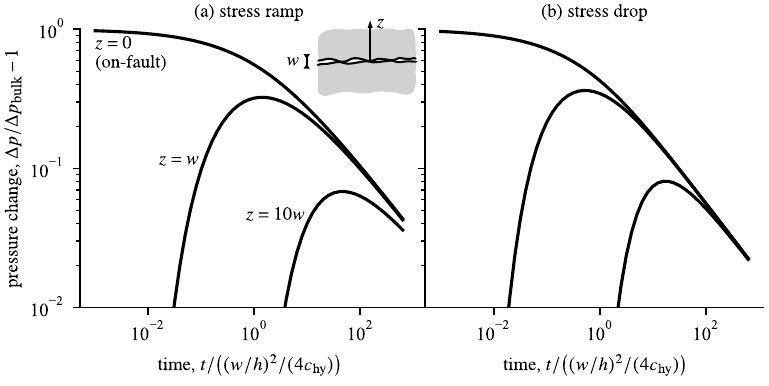}
  \caption{Ratio of on-fault to off-fault pore pressure change ($\Delta p\rightarrow\Delta p_\mathrm{bulk}$ as $z\rightarrow\infty$) in a diffusion model where a fault of width $w$ is contained in a poroelastic bulk under imposed normal stress conditions. We have assumed that the Skempton coefficient of the fault is twice that of the bulk ($B_\mathrm{f}=2B$). (a) Result for a step change in normal stress. (b) Result for a ramp change in normal stress. Parameter $c_\mathrm{hy}$ is the hydraulic diffusivity of the bulk, and $h=S/S_\mathrm{f}$ is the ratio of bulk to fault storage capacity. Solutions are given in Supplementary Materials.}
  \label{fig:diffusion}
\end{figure}

An interesting insight into the drainage state of the fault in our experiments is given by the apparent discrepancy between the pore pressure variations observed during loading and those observed during fast unloading (Figure \ref{fig:dpload}). This limited pore pressure rise during loading is likely due to the high drainage rate compared to the loading rate. This effect can be captured by a simple pore pressure diffusion model of a fracture of width $w$ in an infinite poroelastic medium, subject to a normal stress change (solution details are given in  the Supplementary Materials). In both cases of an instantaneous normal stress change (step) and constant rate (ramp), the pore pressure change within the fault initially reflects the Skempton coefficient of the fault itself, but rapidly diverges and then tracks the pore pressure of the surrounding bulk (Figure \ref{fig:diffusion}). The characteristic timescale for that transition is $t_\mathrm{diff} = (w/h)^2/(4c_\mathrm{hy})$, where $c_\mathrm{hy}$ is the hydraulic diffusivity of the bulk, and $h=S/S_\mathrm{f}$ is the ratio of bulk to fault storage capacity. This characteristic time is the one that appears in the fault recharge problem analysed by \citet{aben23}. Since it is governed by the fault width $w$, it tends to be very short in practice for thin faults: for a diffusivity of the order of $10^{-5}$~m$^2$/s and $h$ of the order of $1$, we obtain a diffusion time of the order of $25$~ms with a fault width of $w=1$~mm, the latter being clearly an upper bound for the actual width of the saw-cut in our experiments.

Therefore, at timescales of the order of seconds (for instance, during static loading), the fault is never truly in an undrained state, and what is detected by our sensors is not purely representative of the fault zone itself due to substantial leak-off. During dynamic events with durations of the order of milliseconds, leak-off is minimal and the pore pressure variations are closer to those of the undrained fault. This analysis is consistent with the systematic results of \citet{rudnicki25} in the context of coupled diffusion with rate-and-state frictional effects.

The static vs. dynamic pore pressure changes shown in Figure \ref{fig:dpload} are consistent with undrained conditions during stick-slip, with Skempton coefficient of the fault close to $1$, and partially drained conditions during static loading, with a bulk Skempton coefficient of the order of $0.5$ or less, consistent with that of intact Westerly granite \citep[e.g.][]{elsigood25}.

\subsection{Implications for stability of slip and fault dynamics}

In our experiments, we observed rapid stick-slip events together with large (dilatant) pore pressure changes: the pore pressure drops did not lead to a systematic stabilisation of slip, as opposed to observations on initially intact rocks \citep{aben21}. It is clear that any potential stabilising effect of dilatancy (poroelastic or inelastic) depends on a number of material parameters and conditions \citep{rudnicki88,segall95}: it is thus likely that our experiments were conducted in a regime where dilational effects are not sufficient to impact the overall stability of slip.

While the role of inelastic dilation on slip stability has been studied systematically in both slip-dependent and rate-and-state friction frameworks \citep{rudnicki88,segall95,mei25}, the role of poroelastic pressure changes due to normal stress variations is less clear. \citet{heimisson21} provided a complete analysis of slip stability for dilatant, rate-and-state faults within poroelastic materials, but their analysis is not easily applicable to our much simpler experimental geometry, more akin to a spring-slider system. Here, we provide a stability analysis of slip for rate-and-state faults with inelastic dilation (similar to the problem treated by \citet{segall95}) that also includes normal stress changes linked to loading geometry (similar to the configuration analysed by \citet{dieterich92b}) and poroelastic pore pressure changes within the fault. We only treat the undrained regime, which yields a simple, interpretable solution, and leave a complete treatment that includes interactions with the poroelastic bulk for future work.

We consider a geometry similar to that analysed by \citet{dieterich92b} (Supplementary Figure S2): a fault oriented at an angle $\psi$ with respect to the maximum principal stress is loaded via a spring of axial stiffness $k$ by a remote displacement $u_0$. The block's axial motion is denoted $u$, and the fault slip is $\delta=u/\cos\psi$. The lateral stress is the confining pressure $P_\mathrm{c}$ and the axial stress is the sum of $P_\mathrm{c}$ and the differential stress $Q$. The geometry implies that the shear stress $\tau$ and normal stress $\sigma_\mathrm{n}$ are
\begin{align}
  \tau &= (Q/2)\sin(2\psi),\\
  \sigma_\mathrm{n} &= P_\mathrm{c} + (Q/2)(1-\cos(2\psi)),
\end{align}
so that
\begin{equation}
  \sigma_\mathrm{n} = P_\mathrm{c} + \tau\tan\psi.
\end{equation}
The axial differential stress is related to the load-point and block displacements via $Q = k(u_0-u)$, which translates to
\begin{equation} \label{eq:tau_el}
  \tau = k_\mathrm{s}(\delta_0 - \delta)
\end{equation}
in terms of shear stress, where $k_\mathrm{s} = k\cos\psi\sin(2\psi)/2$ and $\delta_0=u_0/\cos\psi$.

The shear stress is matched by a frictional strength of the form
\begin{equation}
  \tau_\mathrm{f} = \mu(v,\theta)(\sigma_\mathrm{n} - p),
\end{equation}
where $\mu$ is a friction coefficient that depends on slip rate $v$ and a state variable $\theta$, and $p$ is the pore pressure within the fault. In terms of confining pressure, the frictional strength is given by
\begin{equation} \label{eq:tau_f}
  \tau_\mathrm{f} = \frac{\mu}{1-\mu\tan\psi}(P_\mathrm{c} - p).
\end{equation}

Fluid mass conservation provides an evolution equation for the fluid pressure. In principle, we should consider the fully coupled system of the fault and its poroelastic surrounding bulk \citep[e.g.][]{rudnicki25}, but this is beyond the scope of our simple analysis.  Here, we only aim to gain insights into the role of poroelastic pressure changes \emph{within} the fault on the stability of slip. Therefore, it is sufficient to consider only the undrained limit, neglecting the flux in and out the fault. The pore pressure is thus given by
\begin{equation}\label{eq:dpfu}
  \dot{p} = B_\mathrm{f} \dot{\sigma_\mathrm{n}} - \dot{w}/(Sw),
\end{equation}
where $B_\mathrm{f}$ is the Skempton coefficient of the fault, $\dot{w}$ is the inelastic variation in fault width $w$, and $S$ is the storage capacity of the fault.

To complete the formulation of the problem, we need an explicit form for the constitutive law and for inelastic changes in fault width. Here, we follow \citet{segall95} and use rate-and-state friction with a rate-dependent dilatancy term:
\begin{equation} \label{eq:mu}
  \mu(v,\theta) = \mu_0 + a\ln(v/v_0) + b\ln(\theta/\theta_0),
\end{equation}
where $\mu_0$ is a reference friction coefficient at the reference velocity $v_0$ and state $\theta_0$, and $a$ and $b$ are material constants. The state evolution law is given by either
\begin{equation}\label{eq:dtheta}
\dot\theta = 1 - \theta v/d_\mathrm{c} \quad\text{or}\quad\dot\theta = \theta v/d_\mathrm{c}\ln(\theta v/d_\mathrm{c}),
\end{equation}
where $d_\mathrm{c}$ is constant. Dilatancy is expressed as \citep{segall95}
\begin{equation}\label{eq:dw}
  \dot{w} = v/d_\mathrm{c}(w-w_0-\epsilon\ln(v/v_0)),
\end{equation}
where $w_0$ is the reference fault width and $\epsilon$ is a constant dilatancy coefficient.

Equations \eqref{eq:tau_el}, \eqref{eq:tau_f}, \eqref{eq:dpfu}, \eqref{eq:mu}, \eqref{eq:dtheta} and \eqref{eq:dw} form a nonlinear system of equations for variables $p$, $v$, $\theta$ and $w$. There is a steady-state solution given by $v=v_0$, $\theta=\theta_0=d_\mathrm{c}/v_0$, $w=w_0$ and $p=p_0$. Small perturbations $v_1, \theta_1, w_1, p_1$ around steady-state are governed by the following linearised system:
\begin{align}
  \frac{P_\mathrm{c}-p_0}{(1-\mu_0\tan\psi)^2}\dot\mu_1 - \frac{\mu_0}{1-\mu_0\tan\psi}\dot{p_1} &= -k_\mathrm{s}v_1, \label{eq:kv1}\\
  \dot{p_1} &= -B_\mathrm{f}\tan\psi k_\mathrm{s} v_1 - \dot{w_1}/(Sw_0),\\
  \dot{\mu_1} &= a \frac{\dot{v_1}}{v_0} + b \frac{\dot{\theta_1}}{\theta_0},\\
  \dot{\theta_1} &= -\frac{\theta_1}{\theta_0}-\frac{v_1}{v_0},\\
  \dot{w_1} &= -\frac{v_0}{d_\mathrm{c}}(w_1 - \epsilon v_1/v_0),
\end{align}
where Equation \eqref{eq:kv1} was obtained by equating the time derivative of Equations \eqref{eq:tau_el} and \eqref{eq:tau_f}, neglecting second-order terms. Seeking solutions in the form $v_1 = Ve^{st}$, $\theta_1=\Theta e^{st}$, $w_1=W e^{st}$ and $p_1=P e^{st}$, we find that nontrivial solutions only exist if
\begin{equation} \label{eq:s}
  \sigma_\mathrm{eff,0}as = -k_\mathrm{s}rd_\mathrm{c}\left[1 + \mu_0\tan\psi(B_\mathrm{f}-1)\right] + \frac{rs}{r+s}\left[\sigma_\mathrm{eff,0}b - \frac{\mu_0\epsilon}{S}\right],
\end{equation}
where we used the notation $r=v_0/d_\mathrm{c}$ as in Appendix A of \citet{segall95}, and $\sigma_\mathrm{eff,0}=(P_\mathrm{c}-p_0)/(1-\mu_0\tan\psi)$ is the steady-state effective normal stress. Our Equation \eqref{eq:s} for the perturbation growth rate is in fact the same as Equation A1 of \citet{segall95} for the undrained case (their $c^*\rightarrow 0$), with the spring stiffness modified by the factor $1+\mu_0\tan\psi(B_\mathrm{f}-1)$ to account for the poroelastic effect and the geometrical coupling between shear and normal stress \citep{dieterich92b}.

Therefore, the stability criterion in the undrained regime is
\begin{equation}\label{eq:stability_maintext}
  k_\mathrm{s}\big(1+\mu_0\tan\psi(B_\mathrm{f}-1)\big) > k_\mathrm{crit-undrained} =  \frac{1}{d_\mathrm{c}}\left(\sigma_\mathrm{eff,0}(b-a) - \frac{\epsilon\mu_0}{S}\right).
\end{equation}
The criterion \eqref{eq:stability_maintext} only differs from that established by \citet{segall95} by the factor $1+\mu_0\tan\psi(B_\mathrm{f}-1)$ that multiplies the shear stiffness. This factor captures two competing effects: (1) the geometrical effect of normal stress decrease with ongoing slip, which correspond to the term $(1-\mu_0\tan\psi)$ \citep[as shown by][]{dieterich92b}, and (2) the poroelastic effect which partially counteracts it (term $\mu_0\tan\psi B_\mathrm{f}$). One can verify that setting $B_\mathrm{f}=1$ completely cancels the geometric effect, so that we return to situation where effective normal stress remains nearly constant during slip, and we retrieve the original undrained stability criterion of \citet{segall95}.

Thus, poroelastic variations are further stabilising fault slip, in addition to possible inelastic effects. We anticipate that fluid diffusion would limit the efficiency of such stabilising effects. As pointed earlier, fluid diffusion from thin faults is characterised by short characteristic timescales (scaling with $w^2$), so that saw-cut laboratory faults are less prone to dilatant stabilisation of slip compared to thicker, gouge-filled shear zones. In our experiments, we did not attempt to measure rate-and-state constitutive parameters, so fully evaluating the stability criterion is not possible. However, we infer that our system and fault properties are such that the stabilising effect of poroelasticity and inelastic dilation (as observed in the preslip phase) were not sufficient to prevent stick-slip.

\section{Conclusions}


We conducted laboratory friction experiments under upper crustal conditions in a water-saturated faulted granite. Direct measurements indicate that on-fault pore pressure is impacted by variations in fault normal stress, as well as variations in slip displacement. The effect of normal stress on pore pressure is consistent with the experimental fault having a Skempton coefficient close to unity. During quasistatic loading, fluid diffusion from the fault to the surrounding bulk limits the magnitude of the observed pore pressure change. During stick-slip events, dynamic unloading produces pore pressure changes that are close to normal stress changes.

Superimposed to these first-order variations, we also observe evidence of inelastic dilation in the period immediately preceding slip events. This inelastic dilation is linked to the nucleation phase of the events: strain gauge measurements show a clear correlation between the onset of inhomogeneous slip along the fault and the onset of precursory pore pressure drops. The amount of precursory slip is limited, of the order of a few \textmu{}m to tens of \textmu{}m at most, but is sufficient to produce detectable pore pressure changes of the order of $0.1$ to $1.4$~MPa, commensurate to what was reported by \citet{proctor20}.

The pore pressure drops observed during the main slip events deviate from the anticipated poroelastic change from normal stress variations. In small events (slip less than around $100$~\textmu{}m), we tend to observe larger pore pressure drops, indicating the likely role of coseismic inelastic dilation. In a few events that accumulated larger slip (slip larger than $500$~\textmu{}m), the observed pore pressure drop was somewhat lower than anticipated, which could be linked to either coseismic compaction or thermal pressurisation effects.

The role of poroelastic pore pressure variations in faults sliding under variable normal stress appears to be stabilising: in the limiting case of a Skempton coefficient equal to $1$ and under undrained conditions, all normal stress variations are compensated by pore pressure changes, and the fault operates under constant effective normal stress.

Our experimental results have been obtained on saw-cut, bare-rock surfaces. While poroelastic effects should be a universal feature of slip under variable normal stress, inelastic dilation or compaction should depend critically on the state of the interface and the presence of gouge. \citet{brantut20,aben21} report systematically large dilation caused by slip in initially intact material. \citet{proctor20} show that bare-rock surfaces tend to produce dilation, but that compaction is also possible in the presence of gouge. Shear-induced compaction was also inferred by \citet{faulkner18} in gouge experiments taken to large slip displacements. Transient gouge dilation and pore pressure drops were observed by \citet{affinito25} in response to velocity steps. This variety of observations all point towards the consolidation state of the fault zone material (in the sense of soil mechanics) being the key parameter controlling the evolution of fault zone porosity and of the slip-induced pore pressure changes. Therefore, extrapolation of laboratory observations to nature remains subject to a systematic evaluation of the consolidation state of natural fault rocks, as well as their poroelastic properties.



\paragraph{Acknowledgments} Funding from the UK Natural Environment Research Council (Grant NE/S000852/1 to NB), the European Research
Council under the European Union's Horizon 2020 research and innovation 
programme (project RockDeaf, Grant agreement 804685 to NB; project RockDeath, Grant agreement 101088963 to NB; project Hope, Grant agreement 101041966 to FXP), and a Philip Leverhulme Prize from the Leverhulme Trust (NB), is gratefully acknowledged. Comments by two anonymous reviewers helped clarify the manuscript.


\end{document}


\maketitle

\section*{Solution of diffusion problem}

Consider a fracture of width $w$ in an infinite, isotropic poroelastic medium, subject to a varying normal stress $\sigma_\mathrm{n}$. In the bulk (coordinates $z>0$), pore pressure evolves as
\begin{equation} \label{eq:bulk}
  \frac{\partial p}{\partial t} = c_\mathrm{hy} \frac{\partial^2p}{\partial z^2} + B' \dot{\sigma_\mathrm{n}}, \quad (z>0),
\end{equation}
where $c_\mathrm{hy}=k/(\eta S)$ is the hydraulic diffusivity of the bulk, $k$ is its permeability, $S$ is its storage capacity, $\eta$ is the fluid viscosity and $\dot{\sigma_\mathrm{n}}$ is the rate of change of stress normal to the fault. The coefficient $B'$ is the Skempton coefficient $B$, multiplied by a geometrical factor that relates the change of mean stress in the bulk to the change of stress in the direction normal to the fault. Under triaxial conditions, for a fracture oriented at an angle $\psi$ with respect to the maximum principal stress, that geometrical factor is given by $2/3(1-\cos2\psi)$, i.e., $B' = 2B/3(1-\cos2\psi)$.


The fracture (positioned at $z=0$) acts as a reservoir of volume $w$ per unit area, and the pore pressure there is given by
\begin{equation} \label{eq:fault}
  \frac{\partial p}{\partial t} - \frac{2k}{w\eta S_\mathrm{f}}\frac{\partial p}{\partial z} = B_\mathrm{f}\dot{\sigma_\mathrm{n}}, \quad (z=0),
\end{equation}
where $S_\mathrm{f}$ is the storage capacity of the fracture, and $B_\mathrm{f}$ is its Skempton coefficient (close to one, as discussed in Section 4.1 of main text). The solution of \eqref{eq:bulk} together with \eqref{eq:fault} can be found by the Laplace transform method, and is
\begin{equation} \label{eq:sol_integral}
  p(t,z) = p_0 + \int_0^t B\dot{\sigma_\mathrm{n}}(t') F(t-t',z) dt',
\end{equation}
where $p_0$ is the initial uniform pressure, and $F$ is
\begin{equation} \label{eq:F}
  F(t,z) = 1 + (B_\mathrm{f}/B' - 1)e^{2hz/w +   t/t_\mathrm{diff}}\mathrm{erfc}\left[\frac{z h/w}{\sqrt{t/t_\mathrm{diff}}} + \sqrt{t/t_\mathrm{diff}}\right],
\end{equation}
where $h=S/S_\mathrm{f}$ is the ratio of bulk to fault storage capacity, and $t_\mathrm{diff}=(w/h)^2/(4c_\mathrm{hy})$ is the characteristic diffusion time across the fault.

For a sudden normal stress drop (step change $\Delta\sigma_\mathrm{n}$ in normal stress), the bulk pore pressure change far from the fault is
\begin{equation} \label{eq:dp_bulk_dirac}
  \Delta p_\mathrm{bulk} = B'\Delta\sigma_\mathrm{n},
\end{equation}
and the change at position $z$ is
\begin{equation} \label{eq:dp_fault_dirac}
  \Delta p(t,z) = \Delta p_\mathrm{bulk} F(t,z).
\end{equation}

For a constant rate of normal stress (a stress ramp like that anticipated from the elastic loading stage prior to fault slip), the bulk pore pressure is still given by \eqref{eq:dp_bulk_dirac}, where $\Delta\sigma_\mathrm{n}$ is understood to be constantly increasing with time, and the pore pressure at any $z$ is given by integration of \eqref{eq:sol_integral}. The closed-form solution is
\begin{equation} \label{eq:dp_fault_ramp}
  \Delta p(t,z) = \Delta p_\mathrm{bulk}(t) G(t,z),
\end{equation}
where
\begin{align}
  G(t,z) = 1 + &(B_\mathrm{f}/B' - 1)\left\{ \frac{2}{h}\sqrt{\frac{t/t_\mathrm{diff}}{\pi}}e^{-z^2/4c_\mathrm{hy}t} \right. \nonumber\\
  &\qquad- \frac{1+2hz/w}{h^2}\mathrm{erfc}\left[\frac{z h/w}{\sqrt{t/t_\mathrm{diff}}}\right] \nonumber\\
  &\qquad \left. + \frac{1}{h^2}e^{2hz/w +   t/t_\mathrm{diff}}\mathrm{erfc}\left[\frac{z h/w}{\sqrt{t/t_\mathrm{diff}}} + \sqrt{t/t_\mathrm{diff}}\right]\right\}.
\end{align}

\clearpage
\section*{Supplementary figures}

\begin{figure}[h]
  \centering
  \includegraphics{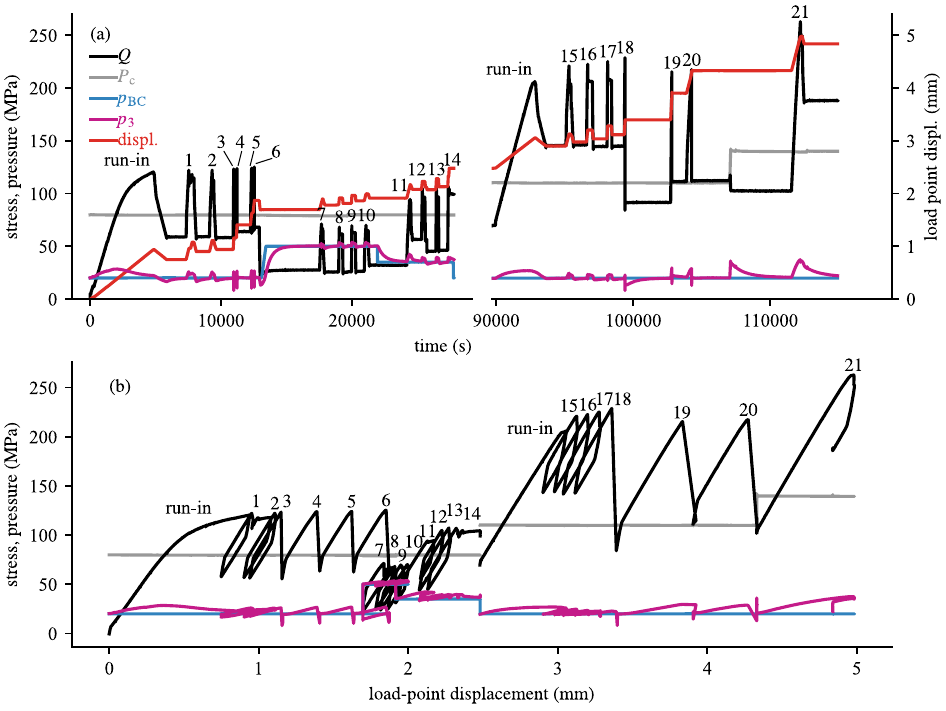}
  \caption{Overview of the experimental sequence. Numbers indicate the stages reported in Table 1 of the main text. (a) Time series. (b) Evolution as a function of load-point displacement. $Q$ denotes the differential stress, $P_\mathrm{c}$ is the confining pressure, $p_\mathrm{BC}$ denotes the pore pressure applied as boundary condition at the sample edge, and $p_3$ is the pore pressure recorded at sensor 3 (on-fault).}
  \label{fig:overview_time_slip_mod}
\end{figure}

\begin{figure}[h]
  \centering
  \includegraphics{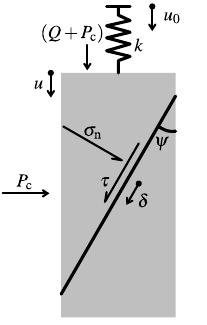}
  \caption{Schematic of the spring-slider geometry employed for the linear stability analysis in Section 4.5 of the main text.}
  \label{fig:stability_sketch}
\end{figure}